\documentstyle[12pt,bezier,epsfig]{article}
\begin{document}
\def \inbar{\vrule height1.5ex width.4pt depth0pt}
\def \xC{\relax\hbox{\kern.25em$\inbar\kern-.3em{\rm C}$}}
\def \xR{\relax{\rm I\kern-.18em R}}
\newcommand{\xZ}{Z \hspace{-.08in}Z}
\newcommand{\xbe}{\begin{equation}}
\newcommand{\xee}{\end{equation}}
\newcommand{\xbea}{\begin{eqnarray}}
\newcommand{\xeea}{\end{eqnarray}}
\newcommand{\xnn}{\nonumber}
\newcommand{\xkt}{\rangle}
\newcommand{\xbr}{\langle}
\newcommand{\xlll}{\left( }
\newcommand{\xrrr}{\right)}
\newcommand{\xcun}{\mbox{\tiny${\cal N}$}}
\title{Two-Component Formulation of the Wheeler-DeWitt~Equation}
\author{Ali Mostafazadeh\thanks{E-mail: alimos@phys.ualberta.ca}\\ \\
Theoretical Physics Institute, University of Alberta, \\
Edmonton, Alberta,  Canada T6G 2J1.}
\date{	}
\maketitle

\begin{abstract}
The Wheeler-DeWitt equation for the minimally coupled
FRW-massive-scalar-field minisuperspace is written as
a two-component Schr\"odinger equation with an explicitly
`time'-dependent Hamiltonian. This reduces the solution
of the Wheeler-DeWitt equation to the eigenvalue problem
for a non-relativistic one-dimensional harmonic oscillator
and an infinite series of  trivial algebraic equations whose
iterative solution is easily found. The solution of these
equations yields a mode expansion of the solution of the
original Wheeler-DeWitt equation. Further analysis of the
mode expansion shows that in general the solutions of the 
Wheeler-DeWitt equation for this model are doubly graded, i.e.,
every solution is a superposition of two definite-parity solutions.
Moreover, it is shown that the mode expansion of both
even and odd-parity solutions is always infinite. It may
be terminated artificially to construct approximate solutions.
This is demonstrated by working out an explicit example which
turns out to satisfy DeWitt's boundary condition at initial
singularity. 
\end{abstract}

\baselineskip= 24pt

\newpage

\section{Introduction}
It is well-known that the Wheeler-DeWitt equation for
the minimally coupled FRW-massive-scalar-field
\footnote{The scalar field is assumed to be
homogeneous.} minisuperspace
can be written in the form of the Klein-Gordon equation
\footnote{This equation corresponds to simplest choice of
factor ordering.} in $(1+1)$ dimensions \cite{page1990},
	\xbe
	\left[ -\frac{\partial^2}{\partial\alpha^2}+
	\frac{\partial^2}{\partial\phi^2}+\kappa\,e^{4\alpha}
	-m^2\,e^{6\alpha}\phi^2\right]\psi(\alpha,\phi)=0\;,
	\label{wdw1}
	\xee
where $\alpha=\ln a$,  $a$ is the scale factor of the FRW
metric
	\xbea
	ds^2&=&-dt^2+a^2(t)\left[ dr^2+S_\kappa(r)(d\theta^2+
	\sin^2\theta \:d\varphi^2)\right]\;,\xnn\\
	S_\kappa&:=&\left\{\begin{array}{cccc}
	\sinh r&{\rm for}&\kappa=-1:&{\rm open~universe}\\
	r&{\rm for}&\kappa=0:&{\rm flat~universe}\\
	\sin r&{\rm for}&\kappa=1:&{\rm closed~universe},
	\end{array}\right.\xnn
	\xeea
and $\phi$ is a real scalar field of mass $m$. It is also
well-known that unlike the massless case \cite{massless} exact
solutions of this equation are not known. The approximate
solutions are usually obtained using the semiclassical (WKB)
methods, \cite{hawking-page}.

Recently, I have used a two-component formulation of the
Klein-Gordon equation to study the geometrical phases
induced on a Klein-Gordon field due to a dynamical
background spacetime geometry and electromagnetic
or scalar interactions, \cite{p19}. In the present article, I shall
try to apply the same approach for the treatment of the
Wheeler-DeWitt equation (\ref{wdw1}).

\section{Two-component Form of the Wheeler-DeWitt Equation}

The idea of the two-component formalism \cite{2-point} is
to transform the Klein-Gordon equation which is of second
order in time-derivative, into a system of two coupled
differential equations which are first order in the
time-derivative. The latter may
be viewed as a Schr\"odinger equation. In this way one can
apply the well-developed techniques of solving the 
Schr\"odinger equation to obtain solutions of the
Klein-Gordon equation.  The application of the two-component
formalism for the Klein-Gordon equation is described in detail in
Ref.~\cite{p19}. Here I only include the results which are
relevant for the solution of Eq.~(\ref{wdw1}).

Eq.~(\ref{wdw1}) can be expressed in the form:
	\xbe
	\left[\eta^{\mu\nu}\frac{\partial }{\partial X^\mu}
	\frac{\partial }{\partial X^\nu}+V(X)\right]\psi=0\;,
	\label{wdw2}
	\xee
where $\mu,\nu=0,1$, $(\eta^{\mu\nu}):={\rm diag}(-1,1)$,
$X^0:=\alpha$, $X^1:=\phi$, and
	\xbe
	V(X):=\kappa\,e^{4X^0}-m^2\,e^{6X^0}(X^1)^2
	=\kappa\,e^{4\alpha}-m^2\,e^{6\alpha}\phi^2=
	\kappa\,a^4-m^2\,a^6\,\phi^2\;.
	\label{V_}
	\xee
Hence, $\alpha$ plays the role of `time.'  Furthermore, defining
	\xbe
	\Psi:=\frac{1}{\sqrt{2}}\left(\begin{array}{c}
	\psi+i\dot\psi\\
	\psi-i\dot\psi\end{array}\right)\;,
	\label{Psi}
	\xee
one can easily check that Eq.~(\ref{wdw2}) is equivalent
to the Schr\"odinger equation
	\xbe
	i\dot\Psi=H\Psi\;,
	\label{sch-eq}
	\xee
with the explicitly `time'-dependent Hamiltonian:
	\xbe
	H=\frac{1}{2}\:\left(
	\begin{array}{cc}
	1+D&-1+D\\
	1-D&-1-D\end{array}\right)\;,
	\label{H}
	\xee
where a dot represents differentiation
with respect to $X^0=\alpha$ and
	\xbe
	D:=-\frac{\partial^2}{\partial (X^1)^2}-V=
	-\frac{\partial^2}{\partial \phi^2}-V\;.
	\label{D}
	\xee

The two-component Hamiltonian (\ref{H}) has the following
remarkable properties:
	\begin{itemize}
	\item[---] It belongs to the (dynamical) Lie algebra
$su_s(1,1)\otimes su_u(1,1)$. Here $su_s(1,1)$ is the
standard representation of $su(1,1)$ with the basis
	\[ K_3=J_3=\frac{1}{2}\:\sigma_3\;,~~~~
	K_a=iJ_a=\frac{i}{2}\:\sigma_a\;,~~~~(a=1,2)\;,\]
where $\sigma_i$ are Pauli matrices, and $su_u(1,1)$
denotes the unitary representation of $su(1,1)$ constructed
out of the position $x$ and momentum $p$ operators of the
one-dimensional quantum mechanics, with the basis
$\{ (x^2-p^2)/4,(x^2+p^2)/4,(xp+px)/4\}$, \cite{jackiw}.
This is easily seen, by writing $H$ in the form
$H=K_-+K_+D$, where $K_\pm:=K_3\pm K_2$,
and observing that the operator $D$ belongs to
$su_u(1,1)$.
	\item[---] It is the square root of the operator $D$,
in the sense that it satisfies
	\xbe
	H^2=\left(
	\begin{array}{cc}
	D&0\\
	0&D\end{array}\right)=D\;.
	\label{H^2}
	\xee
This is reminiscent of  Dirac's treatment of the non-interacting
Klein-Gordon equation in $(3+1)$-dimensional Minkowski space,
where the Dirac operator plays the role of the square root of the
Klein-Gordon operator.
	\end{itemize}

Next let us examine the eigenvalues $E_n$ and eigenvectors
$\Psi_n$ of the Hamiltonian $H$. It is not difficult to show that
due to the particular form of $H$, the eigenvalue equation
	\xbe
	H\Psi_n=E_n\Psi_n\;,
	\label{eg-va}
	\xee
leads to the following relations:
	\xbe
	\Psi_n=\frac{1}{\sqrt{2}}\left(\begin{array}{c}
	1+E_n\\
	1-E_n\end{array}
	\right)\Phi_n\;,~~~~~~
	D\Phi_n=E_n^2\Phi_n\;.
	\label{eg-va-2}
	\xee
Therefore the eigenvalue problem for $H$ is
reduced to that of $D$. Substituting Eq.~(\ref{D}) in the
second equation in (\ref{eg-va-2}), one has:
	\xbe
	\left[-\frac{1}{2}\frac{\partial^2}{\partial\phi^2}+
	\frac{1}{2}\:\omega^2\phi^2\right]\Phi_n={\cal E}_n
	\Phi_n\;,
	\label{eg-va-3}
	\xee
where $\omega:=me^{3\alpha}=ma^3$ and ${\cal E}_n:=
(E_n^2+\kappa e^{4\alpha})/2=(E_n^2+\kappa a^4)/2$.
Eq.~(\ref{eg-va-3}) is the eigenvalue equation for a simple
harmonic oscillator with frequency $\omega$ and unit
mass. Hence:
	\xbea
	{\cal E}_n&=&\omega(n+\frac{1}{2})\:=\:ma^3(n+
	\frac{1}{2})\;,
	\label{cue-n}\\
	\Phi_n&=&N_n H_n(\sqrt{\omega}\phi)\:e^{-\frac{1}{2}
	\omega\phi^2}\:=\:N_nH_n(\sqrt{ma^3} \phi)\:
	e^{-\frac{1}{2}ma^3\phi^2}\;,
	\label{phi-n}
	\xeea
where $H_n$ are Hermite polynomials and $N_n$ are
normalization constants. The latter are
determined by the requirement that $\Phi_n$ be normalized
with the usual $L^2(\xR)$ norm. This leads to 
	\[ N_n=\left[ \frac{\omega}{\pi( 2^n n!)^2}\right]^{1/4}=
	\left[ \frac{m\,a^3}{\pi( 2^n n!)^2}\right]^{1/4}\;.\]
Note that $\Phi_n$ and consequently $\Psi_n$ tend to zero
as $a\to 0$ or $a\to\infty$.

Substituting Eq.~(\ref{cue-n}) in the expression for ${\cal E}_n$,
one has:
	\xbe
	E_{\pm n}=\pm\sqrt{\omega(2n+1)-\kappa e^{4\alpha}}=
	\pm\sqrt{me^{3\alpha}(2n+1)-\kappa e^{4\alpha}}=
	\pm a\sqrt{ma(2n+1)-\kappa a^2}\:.
	\label{e-2}
	\xee
Thus, for $\kappa=-1$ or $0$, $E_{\pm n}$ are real, whereas
for $\kappa=1$ and $a>m$, there are eigenvectors
$\Psi_{\pm n}$ with imaginary eigenvalue. They correspond
to the integers $n$ satisfying $m(2n+1)<a$.
	
So far, I have given the solution for the eigenvalue problem
for the Hamiltonian $H$. The eigenvectors of $H$, however,
do not solve the Schr\"odinger equation. This is mainly
because $H$ is explicitly `time'-dependent. 
The general solution of the Schr\"odinger equation for an
arbitrary explicitly time-dependent Hamiltonian is not known.
However, in the cases where the eigenvalue problem for
the Hamiltonian is solvable, there are techniques based on
the adiabatic approximation \cite{schiff, p16,p18} which
may be used to obtain approximate and occasionally exact
solutions. These techniques can only be employed if
the Hamiltonian is self-adjoint. In order to examine the
self-adjointness of the Hamiltonian $H$, one first needs a
notion of an inner product on the space ${\cal H}$ of 
two-component wave functions $\Psi$. 

\section{Hilbert Space Structure of ${\cal H}$}

As a vector space, the space ${\cal H}$ of  two-component
wave functions is isomorphic to $L^2(\xR) \oplus L^2(\xR)$.
Here by the vector space $L^2(\xR) \oplus L^2(\xR)$, I mean the
vector space inner sum of two copies of $L^2(\xR)$, not the Hilbert
space inner sum of these spaces where the sum is also endowed
with a canonical inner product. Furthermore, in what follows I allow
for negative norms and the term Hilbert space refers to a complete
not necessarily positive-definite inner product space \cite{wald}.
${\cal H}=L^2(\xR) \oplus L^2(\xR)$ may be viewed as a
restriction on the solutions of the Wheeler-DeWitt equation,
since for finite $\alpha$ the elements of $L^2(\xR)\oplus L^2(\xR)$
vanish at `spatial' infinity, i.e., $\phi=\pm\infty$. This choice seems,
however, to be a natural and at the same time a rather general one.
It is only restrictive as far as the boundary condition at `spatial' infinity
is concerned. It is still quite general, for it does not correspond to any
fixed boundary conditions at $\alpha=\pm\infty$. Furthermore,
although ${\cal H}$ has two vector subspaces with $L^2$ inner
products, an inner product on ${\cal H}$ is not a priori fixed.

A Hermitian inner product on ${\cal H}$ may be given by
the inner product on $L^2(\xR)$ and a Hermitian two-by-two
matrix $h$. A particularly advantageous choice for $h$ is
$h=\sigma^3={\rm diag}(1,-1)$, i.e.,
	\xbe
	(\Psi,\tilde\Psi):=\xbr\Psi^1|\tilde\Psi^1\xkt-\xbr\Psi^2|
	\tilde\Psi^2\xkt\;,
	\label{in-pr}
	\xee
where $\xbr~|~\xkt$ denotes the inner product on $L^2(\xR)$.
This choice, on the one hand, renders $H$ self-adjoint.
On the other hand, it reduces to the Klein-Gordon inner
product\footnote{For more details see Ref.~\cite{p19}.}
for the solutions (\ref{Psi}) of the Schr\"odinger
equation (\ref{sch-eq}).

Adopting this inner product on ${\cal H}$, one can compute
	\xbe
	(\Psi_{\pm m},\Psi_{\pm n})=(E_{\pm m}^*+E_{\pm n})
	\delta_{mn}\;.
	\label{eg-in-pr}
	\xee
Here I have used (\ref{eg-va-2}) and the fact that $\Phi_n$
are orthonormal. In particular, $(\Psi_{\pm n},\Psi_{\pm n})=
2{\rm Re}(E_{\pm n})=\pm 2 |{\rm Re}(E_{ n})|$. Hence, the
eigenvectors $\Psi_{\pm n}$ with imaginary eigenvalues (if
they exist) have zero norm; they are null vectors. This is of
course to be expected since the Hamiltonian is
self-adjoint.\footnote{Note that it is the non-positive definiteness
of the corresponding norm that allows for imaginary eigenvalues
and null eigenvectors to exist.} Furthermore, note that the null
eigenvectors are only present for the $\kappa=1$ case.

Another interesting consequence of Eq.~(\ref{eg-in-pr})
is that
	\xbe
	(\Psi_{\pm n},\Psi_{\mp n})=\left\{\begin{array}{cccc}
	0&~~~~~{\rm for}& E_n&{\rm real}\\
	E_{\mp n}&~~~~~{\rm for}& E_n&{\rm imaginary}.
	\end{array}\right.
	\label{decom}
	\xee
Hence, if we view ${\cal H}$ as the linear span of $\Psi_{\pm n}$,
it can be orthogonally decomposed as 
	\xbe
	{\cal H}={\cal H}^{(-)} \oplus {\cal H}^{(0)}
	\oplus {\cal H}^{(+)}\;, 
	\label{H=-0+}
	\xee
where ${\cal H}^{(-)},~{\cal H}^{(0)}$, and ${\cal H}^{(+)}$
are the linear span of the negative-norm, null, and positive-norm
eigenvectors. Note that this decomposition is `time'-dependent (it
depends on $a=e^\alpha$), and not every null vector belongs to
${\cal H}^{(0)}$. Furthermore, Eq.~(\ref{decom}) implies that the
basis $\{\Psi_n\}$ (resp.~$\{\Psi_{-n}\}$) of ${\cal H}^{(+)}$ (resp.\
${\cal H}^{(-)}$) is orthogonal, whereas ${\cal H}_0$ can
be orthogonally decomposed into the sum of  two-dimensional
subspaces ${\cal H}_n^{(0)}={\rm Span}(\Psi_n
\oplus\Psi_{-n})$ with $n\leq (a/m-1)/2$, namely
	\xbe
	{\cal H}^{(0)}=\bigoplus_{n=1}^N {\cal H}_n^{(0)}\;,
	\label{H-0=}
	\xee
where $N:=[(a/m-1)/2]$ is the largest integer part of $(a/m-1)/2$.

\section{Adiabatic Approximation}

Consider the ansatz 
	\xbe
	\Psi=e^{i\zeta_{{\xcun}}}\,\Psi_{\xcun}\;,
	\label{ansatz}
	\xee
for the solution of the Schr\"odinger equation (\ref{sch-eq}), where
$\zeta_{\xcun}$ is a complex-valued function of  $\alpha$,
and ${\xcun}=\pm n \in\xZ$. Plugging (\ref{ansatz}) in (\ref{sch-eq})
and making use of the eigenvalue equation (\ref{eg-va})  and
the form (\ref{eg-va-2}) of the eigenvectors $\Psi_{\xcun}$, one
has:
	\xbea
	\left[ -(\dot\zeta_{\xcun}+E_{\xcun})(1+E_{\xcun})+
	i\dot E_{\xcun}\right]\Phi_n+i(1+E_{\xcun})\dot\Phi_n&=&0
	\;,\xnn\\
	\left[ -(\dot\zeta_{\xcun}+E_{\xcun})(1-E_{\xcun})-
	i\dot E_{\xcun}\right]\Phi_n+i(1-E_{\xcun})\dot\Phi_n&=&0
	\;,\xnn
	\xeea
Adding and subtracting both sides of these equations, one
finds:
	\xbe
	(\dot\zeta_{\xcun}+E_{\xcun})\Phi_n-i\dot\Phi_n=0\;,~~~~~
	\dot E_{\xcun}=0\;.
	\label{ad-ap}
	\xee
Here the first equation implies that  for $\tilde n\neq n$, $A_{
\tilde n n}:=i\xbr\Phi_{\tilde n}|\dot\Phi_n\xkt=0$. This
requirement together with
the second relation in (\ref{ad-ap}) signify the conditions
under which the adiabatic approximation would yield an exact
solution of the Schr\"odinger equation. Clearly, $\dot E_{\xcun}
\neq 0$ for the case under consideration. Thus the adiabatic
approximation cannot be exact. However, approximate solutions
may be sought if $\dot E_{\xcun}$ and $A_{\tilde nn}$ are small.

Performing the necessary calculations, one  can show that
for arbitrary non-negative integers $\tilde n$ and $n$:
	\xbea
	A_{\tilde nn}&:=&i\xbr\Phi_{\tilde n}|\dot\Phi_n\xkt\:=\:
	\frac{3i}{4}\,
	\left[\sqrt{n(n-1)}\:\delta_{\tilde n,n-2}-
	\sqrt{(\tilde n)(\tilde n-1)}\:\delta_{\tilde n,n+2}\right]\;,
	\label{A}\\
	\dot E_{\xcun}&=&\pm\frac{3\left[n+\frac{1}{2}-
	\frac{2\kappa a}{3m}\right]\sqrt{m}\,a^{3/2}}{
	\sqrt{2n+1-\frac{\kappa a}{m}}}\;,
	~~~~{\rm where}~~~~{\xcun}=\pm n\;.
	\label{dot-E}
	\xeea
Eq.~(\ref{A}) indicates that the adiabatic approximation
is not valid, i.e., the solutions of the form (\ref{ansatz})
do not exist. 	
	
Another consequence of Eq.~(\ref{A}) is that in a mode
expansion of a solution $\Psi$ of Eq.~(\ref{sch-eq}), the
even modes $\Psi_{2\xcun}$ and the odd  modes
$\Psi_{2\xcun+1}$ do not mix in the course of evolution.
This suggests the existence of a double grading of the
set of solutions. In particular one has the {\em even}
and {\em odd} solutions:
	\xbe
	\Psi_{\rm even}=\sum_{\xcun=-\infty}^\infty Z_{2\xcun}
	\Psi_{2\xcun}\;,~~~~~
	\Psi_{\rm odd}=\sum_{\xcun=-\infty}^\infty Z_{2\xcun+1}
	\Psi_{2\xcun+1}\;.
	\label{grading}
	\xee

\section{Mode Expansion of the Solutions}

Consider expressing the solution $\Psi$ of the two-component
Wheeler-DeWitt equation~(\ref{sch-eq})
as a linear combination of the eigenfunctions $\Psi_{\xcun}$ of the
Hamiltonian:
	\xbe
	\Psi=\sum_{\xcun=-\infty}^\infty Z_{\xcun}\,\Psi_{\xcun}\;,
	~~~~~Z_{\xcun}=Z_{\xcun}(\alpha)\;.
	\label{mode-exp}
	\xee
Substituting this equation in (\ref{sch-eq}) and using
Eqs.~(\ref{eg-va}) and (\ref{eg-va-2})
to simplify the result, one finds
	\xbea
	\sum_{n=0}^\infty \left[ (\dot\sigma_n+i\sigma_{-n}E_n)
	\Phi_n+\sigma_n\dot\Phi_n\right]&=&0\;,
	\label{I}\\
	\sum_{n=0}^\infty \left[ (\dot\sigma_{-n}E_n+i\sigma_n
	E_n^2+\sigma_{-n}\dot E_n)\Phi_n+\sigma_{-n}E_n
	\dot\Phi_n\right]&=&0\;,
	\label{II}
	\xeea
where
	\[
	\sigma_{\pm n}=\sigma_{\pm n}(\alpha)
	:=Z_n\pm Z_{-n}\;,~~~~~~n=0,1,2,\cdots\;.\]
Note that in general $Z_{-0}\neq Z_0$ and $\sigma_{-0}\neq
\sigma_0$.

Next take the inner product of both sides of Eqs.~(\ref{I})
and (\ref{II}) with $\Phi_{\tilde n}$. Making use of
Eq.~(\ref{A}) and orthonormality of $\Phi_n$, one has:
	\xbea
	&&\dot\sigma_n+iE_n\sigma_{-n}+\frac{3}{4}\,
	\sqrt{(n+1)(n+2)}\,\sigma_{n+2}-\frac{3}{4}\,
	\sqrt{n(n-1)}\,\sigma_{n-2}=0\;,
	\label{III}\\
	&&E_n\dot\sigma_{-n}+iE_n^2\sigma_n+\dot E_n
	\sigma_{-n}+\frac{3}{4}\,\sqrt{(n+1)(n+2)}\,E_{n+2}
	\sigma_{-(n+2)}-\frac{3}{4}\,\sqrt{n(n-1)}\,E_{n-2}
	\sigma_{-(n-2)}=0,~~~~~
	\label{IV}
	\xeea
where $\sigma_{-(0)}=\sigma_{-0}$.

In order to analyze these equations further, let us solve
for $\sigma_{-n}$ from (\ref{III}) and substitute the result in
(\ref{IV}). Doing the same for $\sigma_{-(n\pm 2)}$, one
obtains:
	\xbea
	&&\sigma_{-n}=\frac{i}{E_n}\left[\dot\sigma_n+
	\frac{3}{4}\,\sqrt{(n+1)(n+2)}\,\sigma_{n+2}-\frac{3}{4}\,
	\sqrt{n(n-1)}\,\sigma_{n-2}\right]\;,
	\label{V}\\
	&&\ddot\sigma_n+[E_n^2-\frac{3}{2}(n^2+n+1)]
	\sigma_n+\frac{3}{2}\,\sqrt{(n+1)(n+2)}\,
	\dot\sigma_{n+2}\xnn\\
	&&\hspace{.5cm}
	-\frac{3}{2}\,\sqrt{n(n-1)}\,
	\dot\sigma_{n-2}+(\frac{3}{4})^2\sqrt{(n+1)(n+2)(n+3)(n+4)}\,
	\sigma_{n+4}+\xnn\\
	&&\hspace{.5cm}
	(\frac{3}{4})^2
	\sqrt{n(n-1)(n-2)(n-3)}\,\sigma_{n-4}=0\;.
	\label{VI}
	\xeea
Note that Eq.~(\ref{VI}) links only $\sigma_{2n}$'s,
with $n=0,1,\cdots$, thus making the analysis of
the coefficients of non-negative modes independent
of those of the negative modes.

Eqs.~(\ref{V}) and (\ref{VI}) clearly display the splitting
of the solutions of the Wheeler-DeWitt equation into
even and odd solutions. Since these equations are
linear and there is no correlation between the even
and odd solutions,  every solution is a linear combination
of an even and an odd solution. Hence without loss of
generality, one can restrict to the analysis of
$\Psi_{\rm even}$ and $\Psi_{\rm odd}$.

Before engaging in this analysis, however, I would like
to note that the mode expansion (\ref{mode-exp}) for the
two-component wave function $\Psi$, translates into
	\xbe
	\psi=\sum_{n=0}^\infty\sigma_n\Phi_n\;,
	\label{mode-exp-1}
	\xee
for the one-component wave function. Thus, it is the
coefficients of the zero and positive modes that determine
the solution of the Wheeler-DeWitt equation (\ref{wdw1}).
The negative modes are related to the derivative of
$\psi$ with respect to $\alpha$. Hence, they are also
uniquely determined once the zero and positive modes
are given.

In the following sections, I shall show that the mode expansion
of the solution of the Wheeler-DeWitt equation (\ref{wdw1}) is
always infinite. This is another way of demonstrating the fact
that the adiabatic approximation, which corresponds to including
only a single mode in this expansion, always fails. Nevertheless,
if one tries to blindly apply the adiabatic approximation or a slightly
improved version of it in which one sets all $\sigma_{\pm n}$ to zero,
except  $\sigma_{\pm n_*}$ for some $n_*$, then Eqs.~(\ref{III}) and
(\ref{IV}) yield $\ddot\sigma_{n^*}+E_n^2\sigma_{n^*}=0$. This
corresponds to the eigenvalue equation for a one-dimensional
quantum particle of mass $1/2$ and potential ${\cal V}(\alpha):=
-E_n^2=\kappa e^{4\alpha}-(2n+1)me^{3\alpha}$. Specifically,
$\sigma_{n^*}$ is the zero-energy eigenfunction. For $\kappa=1$,
${\cal V}$ vanishes at $\alpha= -\infty$. It decreases to attain
a negative minimum as $\alpha$ increases, and then monotonically
increases to $+\infty$ as $\alpha\to +\infty$. Hence, if one requires
the zero-energy eigenfunction to vanish at $\alpha=+\infty$ where
${\cal V}=+\infty$, then the mass $m$ must satisfy a peculiar
quantization condition. This is analogous to the mass quantization
advocated in Ref.~\cite{kiefer-88}. However, note that the basic
hypothesis of this argument, namely the validity of the adiabatic
approximation or alternatively the presence of only a single mode
in the mode expansion (\ref{mode-exp-1}), is never satisfied. Hence,
the argument leading to the quantization of mass does not apply.

\section{Even Solutions}

Let us first prove:
	\begin{itemize}
	\item[~] Lemma~1:~{\em An even solution $\Psi_{\rm even}
	=\sum_{\xcun}Z_{2\xcun}\Psi_{2\xcun}$
	is nontrivial, i.e., $\Psi_{\rm even}\neq 0$, if and only if 
	$\sigma_{\pm 0}\neq 0$.}
	\end{itemize}
In order to prove this statement, one sets $n=0$ in
Eqs.~(\ref{III}) and (\ref{IV}). This leads to
	\xbea
	\dot\sigma_0+iE_0\sigma_{-0}+\frac{3\sqrt{2}}{4}\,
	\sigma_2&=&0\;,
	\label{III'}\\
	E_0	\dot\sigma_{-0}+\dot E_0\sigma_{-0}+
	iE_0^2\sigma_{0}+
	\frac{3\sqrt{2}}{4}\,E_2\sigma_{-2}&=&0\;.
	\label{IV'}
	\xeea
Therefore, if $\sigma_{\pm 0}=0$, both $\sigma_2$ and
$\sigma_{-2}$ must vanish.  However, by virtue of
Eqs.~(\ref{VI}) and (\ref{V}), this is sufficient to
conclude that $\sigma_{2\xcun}=0$, for all $\xcun\in
\xZ$. This completes the proof of Lemma~1.

Furthermore, one can prove
	\begin{itemize}
	\item[~] Lemma~2: {\em Every nontrivial even
	solution has an infinite mode expansion.}
	\end{itemize}
The even solutions may nevertheless be constructed
inductively in terms of $\sigma_{\pm 0}$.  This is done by
noting that  Eqs.~(\ref{V}) and (\ref{VI}) yield
trivial algebraic equations for $\sigma_{\pm 2(n+1)}$ in
terms of $\sigma_{\pm 2n}$. Performing an induction
on $n$, one can express the coefficients of all
the modes as  linear combinations of $\sigma_{\pm 0}$ and
its derivatives. For example, one has
	\xbea
	\sigma_2&=&-\frac{1}{\sqrt{2}}(\frac{4}{3})\, 
	(\dot\sigma_0+iE_0\sigma_{-0})\;,
	\label{s2=}\\
	\sigma_4&=&\frac{1}{\sqrt{4!}}(\frac{4}{3})^2\left[
	\ddot\sigma_0+2iE_0\dot\sigma_{-0}+
	(\frac{3}{2}-E_0^2)\sigma_0
	+2i\dot E_0\sigma_{-0}\right]\;.
	\label{s4=}
	\xeea
It is easily seen that as a result of this construction
$\sigma_{2n}$ involves the first $n$ derivatives of
$\sigma_0$, first  $n-1$ derivatives of $\sigma_{-0}$,
and the first $n-1-j$ derivatives of $E_{2j}$, for $j=0,1,
\cdots,n-1$. 

Since according to Eq.~(\ref{dot-E}), $\dot E_n$
diverges at the zero of $E_n$, demanding regularity of
$\sigma_n$ would impose certain conditions on
$\sigma_{\pm 0}$. In fact, in view of the appearance of
$E_n^2$ in Eq.~(\ref{VI}), which is analytic for every
$\alpha$, one can easily show that the only source
of possible singularity is the term proportional to $E_0$
in the expression for $\sigma_2$. Thus, in order to ensure
the regularity of  $\sigma_{2n}$, one must require
that at the zero of $E_0$, i.e.,  $\alpha=\ln(
m/\kappa)$,~ $\sigma_{-0} E_0$ and all its derivatives
must vanish. In other words there must be an open
neighborhood ${\cal O}$ of $\alpha=\ln(m/\kappa)$, in which
$\sigma_{-0}(\alpha)$ is of the form $E_0(\alpha)f(\alpha)$
for some smooth ($C^\infty$) function $f $. In particular
$\sigma_{-0}(\ln(m/\kappa))$ must vanish. Note that for the
flat ($\kappa=0$) and open ($\kappa=-1$) cases, this
condition is void and $\sigma_{2n}$ are regular if
and only if $\sigma_{\pm 0}$ are regular.

Finally note that  termination of the mode expansion
for a nontrivial even solution is equivalent to vanishing
of at least three consecutive $\sigma_{2n}$'s. This would
impose three functional conditions on $\sigma_{\pm 0}$ which
are in general incompatible. This proves Lemma~2.

\section{Odd Solutions}

The structure of the odd solutions is similar to that of the even
solutions. The role of $\sigma_{\pm 0}$ is played by
$\sigma_{\pm1}$. In particular, one has
	\begin{itemize}
	\item[]Lemma~3: {\em An odd solution
	$\Psi_{\rm odd}=\sum_{\xcun} Z_{2\xcun+1} 
	\Psi_{2\xcun+1}$ is
	nontrivial, if and only if  $\sigma_{\pm1}:=Z_1
	\pm Z_{-1}\neq 0$.}
	\end{itemize}
This is a direct consequence of Eqs.~(\ref{III}) and
(\ref{VI}) which also provide an inductive
construction of  the mode coefficients $\sigma_{\pm(
2n+1)}$ in terms of $\sigma_{\pm 1}$. For example
one has:
	\xbea
	\sigma_3&=&-\frac{1}{\sqrt{3!}}(\frac{4}{3}) (\dot
	\sigma_{1}+iE_1\sigma_{-1})\;,
	\label{s3}\\
	\sigma_{5}&=&\frac{1}{\sqrt{5!}}(\frac{4}{3})^2\left[
	\ddot\sigma_1+2iE_1\dot\sigma_{-1}+
	(\frac{9}{2}-E_1^2)\sigma_1+2i\dot E_1
	\sigma_{-1}\right]\;.
	\label{s5}
	\xeea
A natural consequence of this inductive construction
is 
	\begin{itemize}
	\item[]Lemma~4: {\em  Every nontrivial odd
	solution has an infinite mode expansion.}
	\end{itemize}

The regularity condition on the odd solutions
is analogous to the even case. One can again
show that the only source of possible singularity
is the term $E_1\sigma_{-1}$ in the expression for
$\sigma_3$. Hence the regularity of $\sigma_{2n+1}$
and therefore $\psi$ is ensured by requiring that
in the vicinity of the zero of $E_1$, namely
$\alpha=\ln(3m/\kappa)$, $\sigma_{-1}$ must be of
the form $E_1(\alpha) g(\alpha)$, for some smooth
function $g$.

The occurrence of two free functions of $\alpha$ in
the construction of even and odd solutions is in complete
agreement with the general structure of the 
Wheeler-DeWitt/Klein-Gordon equation (\ref{wdw1}) whose
solutions depend
on two boundary or two initial conditions. The latter are given
by two functions of $\alpha$ which in the above procedure
can be used to determine $\sigma_{\pm 0}$ or $\sigma_{\pm 1}$
depending on whether  the solution is even or odd. In general
the solution will be the sum of an even and an odd part. In
this case the boundary (initial) conditions would determine
both $\sigma_{\pm 0}$ and $\sigma_{\pm 1}$. Once these
functions are determined the above iterative construction
yields the mode expansion of the general solution.

\section{Real Solutions}

Probably the best-known proposal for the solution of the
Wheeler-DeWitt equation is that of Hartle and Hawking
\cite{ha-ha}  who suggested the path integral expression
 $\psi \propto \int\exp(-I)$ for the ground state wave
function. Here $I$ stands for the Euclidean action.
Hartle-Hawking wave function is by construction real. 

It is not difficult to see that  the reality
of the one-component solution of the Wheeler-DeWitt
equation implies the reality of the coefficient functions
$\sigma_n$ in its mode expansion (\ref{mode-exp-1}).
This is because $\Phi_n$ are already real.  In view of
this observation and Eq.~(\ref{III}), $E_n\sigma_{-n}$
must be imaginary. However, as discussed in section~2,
for the open and flat universes ($\kappa=-1,0$), $E_n$
is always real so that $\sigma_{-n}$ must be imaginary.
For the close universe ($\kappa=1$)
$E_n$ is real for $a\leq m(2n+1)$ and imaginary
otherwise. Hence in this case ($\kappa=1$)
the opposite holds for $\sigma_{-n}$, i.e.,
it is imaginary for $a< m(2n+1)$ and real for $a> m(2n+1)$. 

These conditions can also be stated as the reality
condition on $\sigma_0$, $\sigma_1$, $\sigma_2$, and
$\sigma_3$ which also generate the solutions. Therefore,
they are rather trivial.

\section{Approximate Solutions}

As shown in the preceding sections, the mode expansion of the
solutions never terminates.  Therefore, one cannot use it to
construct  an exact solution unless one can actually find the
generic expression for the mode coefficients and sum the series.
This is an almost impossible task. However,  one can terminate the
mode expansion artificially and obtain an infinite class of approximate
solutions, $\psi\approx \psi_{n_*}:=\sum_{n=0}^{n_*}\sigma_n\Phi_n$.
The domain of reliability of this solutions will in general depend on the
boundary (initial) conditions, alternatively the functions
$\sigma_{\pm 0}$ and $\sigma_{\pm 1}$.

In this section, I shall examine the simple choice $\sigma_{-0}=
\sigma_{\pm 1}=0$ and $\sigma_0=1$.  Clearly, this corresponds
to a real even solution whose mode coefficients can be obtained
using  Eq.~(\ref{VI}). For example for the closed FRW case where
$\kappa=1$, one has
	\xbea
	\sigma_2&=&0\;,~~~\sigma_4=(\frac{2}{3})^{5/2}(\frac{3}{2}+
	a^4-ma^3)\;,~~~\sigma_6=(\frac{32a^3}{81\sqrt{5}})(3m-4a)\;,\xnn\\
	\sigma_8&=&\frac{4}{243}\sqrt{\frac{2}{35}}(4a^8-40ma^7+
	36 m^2a^6+324a^4-288ma^3+189)\;,\xnn\\
	\sigma_{10}&=&-\frac{16}{10935\sqrt{7}}(96 a^5-1000 ma^4+
	744m^2a^3+3604a-2655m)a^3\;.\xnn
	\xeea
Substituting these relations in the expression for $\psi_n$ and making
use of  the identity \cite{g-r}
	\[ H_n(x)=2^nx^n+\sum_{i=1}^{[n/2]}
	 (-1)^i2^{n-i}(2i-1)!!\left(\begin{array}{c}n\\2i
	\end{array}\right)x^{n-2i}\;,\]
to calculate the mode functions $\Phi_n$, one can compute $\psi_0
=\psi_2,~\psi_4,~\psi_6,~\psi_8$ and $\psi_{10}$. Figures~1 to~5
illustrate three-dimensional plots of these functions. Figures~6 to~9
are plots of the restriction of $\psi_6,~\psi_8$ and $\psi_{10}$ to the
lines $a=0.1,1,10,100$. These figures show that for small $a$ the
finite mode approximation is quite reliable. In fact, as depicted in Fig.~6,
for $a=0.1$, the graphs of $\psi_8$ and $\psi_{10}$ coincide. For
larger $a$ one may improve the approximation by including the
contribution of higher modes. 

Note also that initially $\psi_n$ are non-zero for a
wide range of values of $\phi$. However, as $a$ increases the width
of this region, which is symmetric with respect to the line $\phi=0$,
shrinks. This is easily seen from Figures~6 to~9. Note also that for
large $a$, the solutions picks about $\phi=0$ and the approximate
solutions become less reliable. As $a\to\infty$, all the approximate
solutions seem to vanish except at $\phi=0$.

\section{Discussion and Conclusion}

Probably the most important aspect of the two-component
formulation of the Wheeler-DeWitt equation is the fact that
the construction of the solutions in this formulation does not,
a priori, require fixing a particular inner product or boundary
conditions.  The freedom in the choice of the inner product is
reflected as the independence of the mode expansion from the
choice of the Hermitian matrix $h$ which together with the $L^2$
inner product on $L^2(\xR)$ determines the inner product on
${\cal H}=L^2(\xR)\oplus L^2(\xR)$. As far as the boundary
conditions are concerned, besides the implicit restriction of
vanishing boundary condition at `spatial'  infinity, the analysis
presented here applies generally.

An interesting consequence of the application of the two-component
formalism is a specific double grading of the solutions of the
Wheeler-DeWitt equation. This allows one to write down the
most general solution of this equation as a sum of definite-parity
(even and odd) solutions whose structure can be easily understood.
The two definite-parity parts of the solution do not mix during the
evolution. In particular, if the initial wave function has definite parity,
its parity remains intact throughout the evolution. 

Unfortunately, the mode expansion of the solutions never
terminates.  Thus, a closed expression for an exact solution
cannot be found in this way. However, approximate solutions
can be constructed by artificially terminating the mode
expansion. The domain of applicability of such approximations
depend generally on the boundary (initial) conditions.

An alternative approximation is the adiabatic
approximation which is, however, not valid. The ansatz
of the adiabatic approximation, i.e., $\psi\approx
\sigma_n(\alpha)\Phi_n(\alpha,\phi)$, is reminiscent of the
application of the Born-Oppenheimer approximation in
quantum cosmology originally advocated by Banks \cite{b-o}.
In this approach one views the gravitational
degrees of freedom as the `slow' variables and the matter
degrees of freedom as the `fast' variables and  considers
solutions of the form $\psi=\psi_{\rm gravity}(\alpha)
\psi_{\rm matter}(\alpha,\phi)$. The results of this article
indicate that for $\psi_{\rm matter}=\Phi_n$ such an
approximation cannot be valid.

Another notable implication of the results of this
article is regarding the implicit boundary condition
at the initial singularity $a=0$. If one assumes that the
coefficient functions $\sigma_n$ are regular at
$a=0$, then as a consequence of the appearance
of  $a^{3/4}$ in the expression for $\Phi_n$,
the one-component wave-function $\psi$ vanishes
at $a=0$. This is much in the spirit of an argument
of  DeWitt \cite{bd-1967-I}. Here it follows as a direct
consequence of a regularity condition (which may or may
not hold). As seen from Figures~1-5, the choice
$\sigma_{-0}=\sigma_{\pm 1}=0$ and $\sigma_0=1$
leads to approximate solutions which are regular at
$a=0$. They clearly satisfy DeWitt's boundary conditions.

A problem which is not addressed in this article is that
of the convergence of the mode expansion in ${\cal H}$.
This is a difficult problem and I have not been able
to find the necessary and sufficient conditions under which
the mode expansion converges. In fact, a rigorous convergence
analysis requires a choice of a positive-definite inner product
(norm) on ${\cal H}$. However, as demonstrated in the preceding
section, there are regions of the minisuperspace in which
the contribution from the higher order modes is small.  This may
be viewed as an indication of  possible pointwise asymptotic
convergence of the mode expansion in this regions. This
argument is however far from being satisfactory and further
investigation of this point is necessary.

Finally, I wish to emphasize the following points:
	\begin{itemize}
	\item[---] The fact that the two-component Hamiltonian
(\ref{H}) belongs to the dynamical Lie algebra $su_s(1,1)
\otimes su_u(1,1)$ suggests a further investigation of the
solution of the corresponding Schr\"odinger equation using
the group theoretical methods such as those employed in
solving the Schr\"odinger equation for time-dependent
harmonic oscillators, \cite{oscillator}.  Recall that the
Hamiltonian for a time-dependent harmonic oscillator belongs
to $su_u(1,1)$, \cite{jackiw}. 
	\item[---] The two-component approach to Wheeler-DeWitt
equation has a wider domain of applicability than the simple
minisuperspace model studied in this article. For example, the
addition of an interaction potential for the scalar field would only
complicate the defining equation for $\Phi_n$ and the inductive
construction of the mode expansion. The general structure of the
method would however remain intact. In particular,  the
$\phi$-dependence of the solutions would still be given by $\Phi_n$.
More generally, for any other minisuperspace, in the gauge in which
the shift vector corresponding to DeWitt's supermetric \cite{bd-1967-I}
vanishes, the two-component Hamiltonian  is of the form (\ref{H}),
where $D$ is some second order elliptic operator. Therefore, even
in this case the developments reported in this article are
essentially applicable.
	\end{itemize}

\vspace{.5cm}

{\large{\bf Note:}} After the completion of this work, Refs.\
\cite{kim, dereli}  were brought to my attention where the
authors use two different two-component formulations of the
Wheeler-DeWitt equation. The methods described in these
articles differ substantially from the present article's. I have
also noticed a recent paper by Embacher \cite{embacher}
who defines the operator $D$ of Eq.~(\ref{D}) as the (matter)
energy operator and obtains the mode expansion (\ref{mode-exp-1})
in the context of  the one-component Wheeler-DeWitt equation. 
He does not, however, obtain the equations satisfied by the
mode coefficients. Instead, he focuses on the semiclassical and
adiabatic approximations. The two-component approach
developed in this paper justifies the choice of the mode functions
made by Embacher in a very natural way. It also makes it possible
to obtain the structure of the mode expansion by iteratively
solving for the mode coefficients.

\section*{Acknowledgements}
I would like to thank B.~Darian,  M.~Razavi, and especially
D.~Page for invaluable comments and suggestions. I would
also like to acknowledge the financial support of the Killam
Foundation of Canada.

{
}

\newpage

\begin{figure}
\epsffile{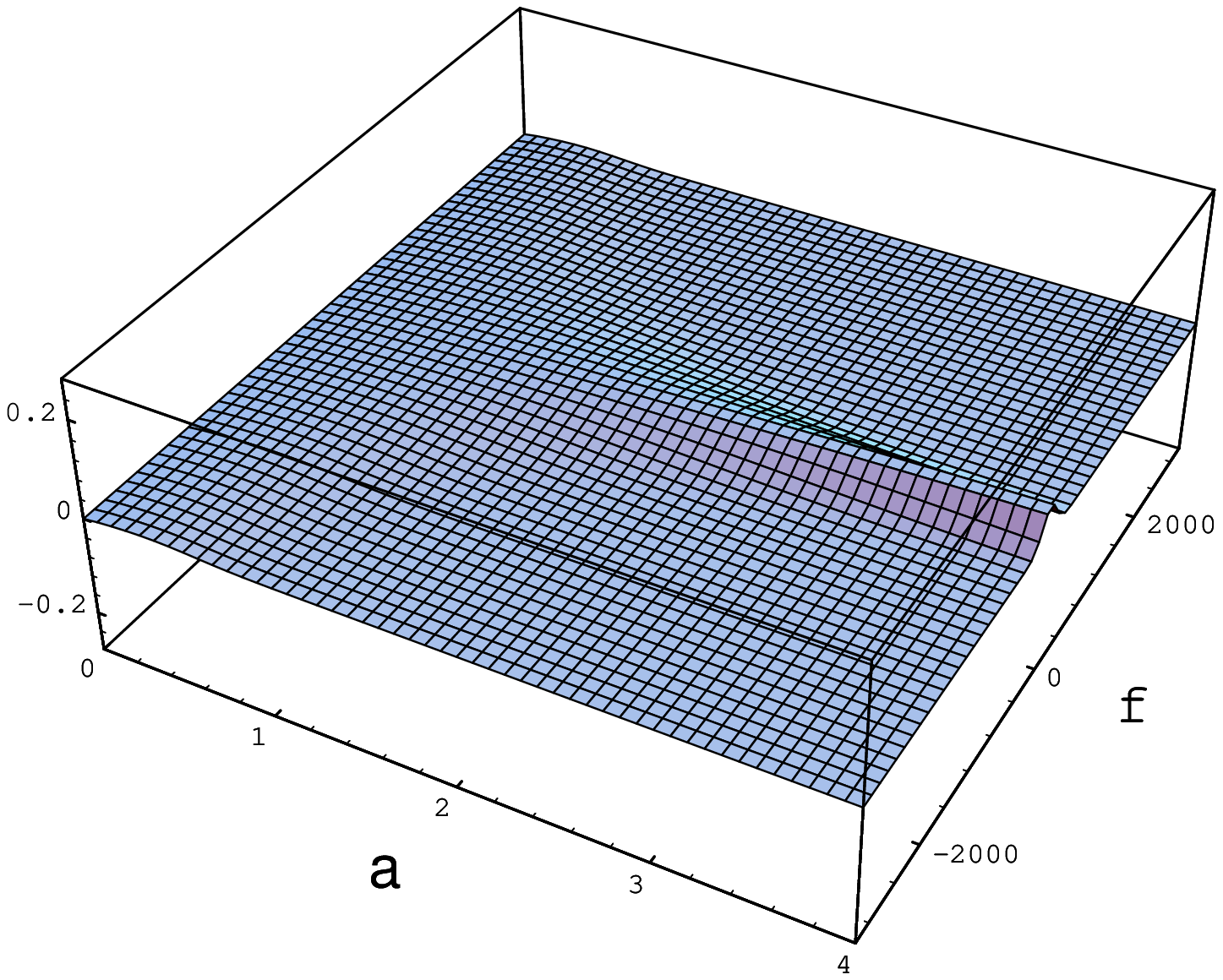}
\caption{Plot of $\psi_0=\psi_2=\sum_{n=0}^2\sigma_n\Phi_n$, for $\sigma_{\pm1}=\sigma_{-0}=0,~\sigma_0=1$ and $m=10^{-6}$.}
\end{figure}

\begin{figure}
\epsffile{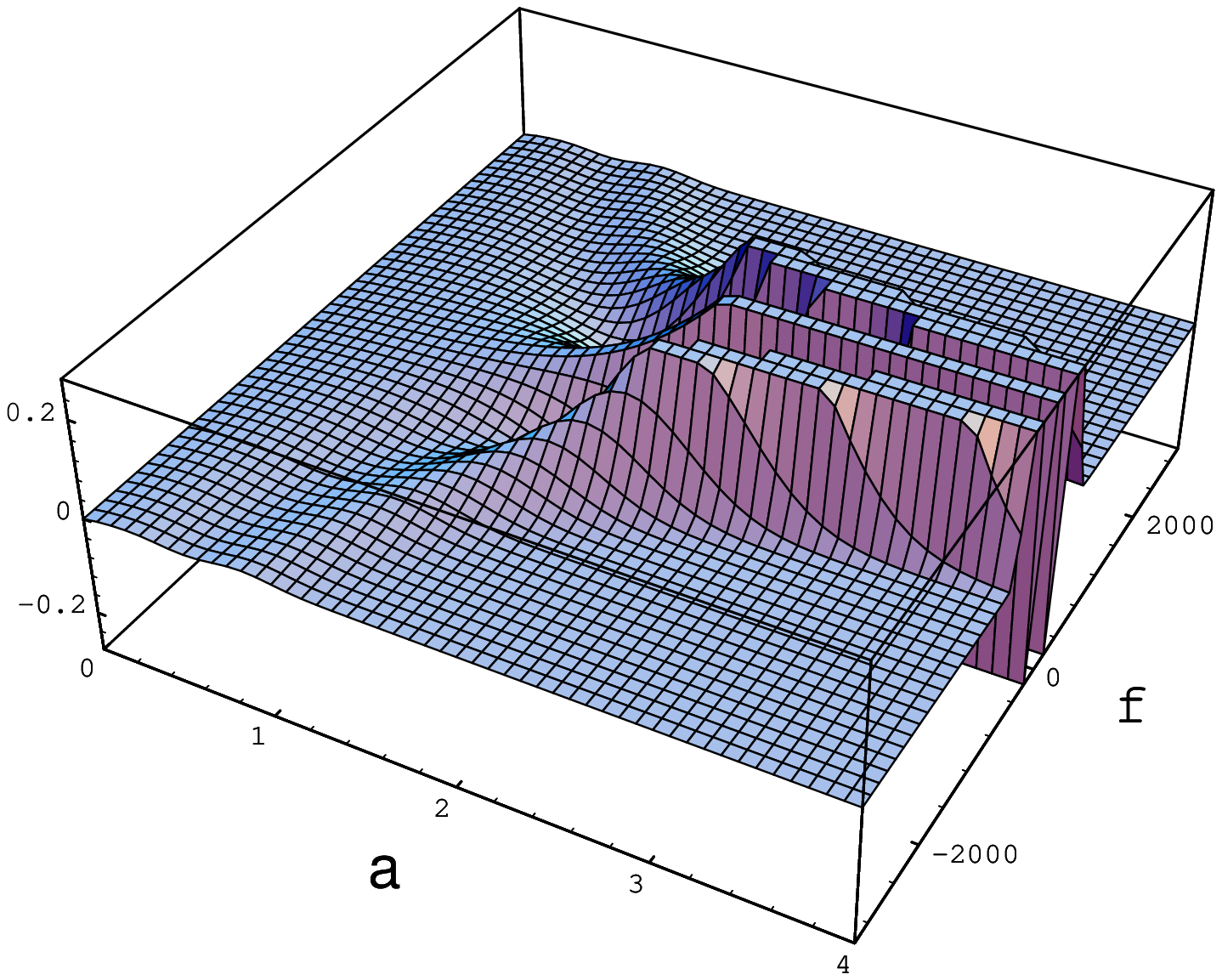}
\caption{Plot of $\psi_4:=\sum_{n=0}^4\sigma_n\Phi_n$, for
$\sigma_{\pm1}=\sigma_{-0}=0,~\sigma_0=1$ and $m=10^{-6}$.}
\end{figure}

\begin{figure}
\epsffile{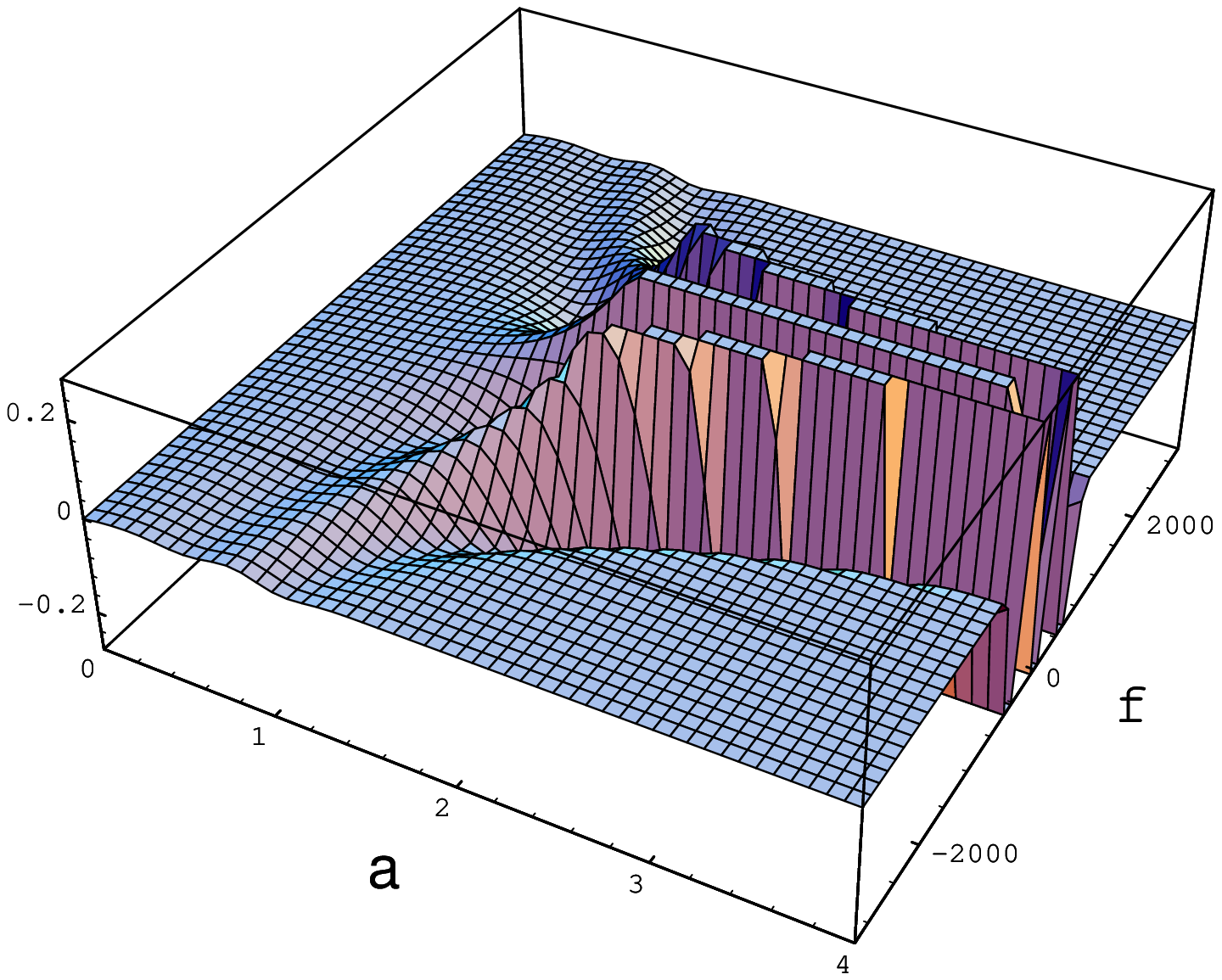}
\caption{Plot of $\psi_6:=\sum_{n=0}^6\sigma_n\Phi_n$, for
$\sigma_{\pm1}=\sigma_{-0}=0,~\sigma_0=1$ and $m=10^{-6}$.}
\end{figure}

\begin{figure}
\epsffile{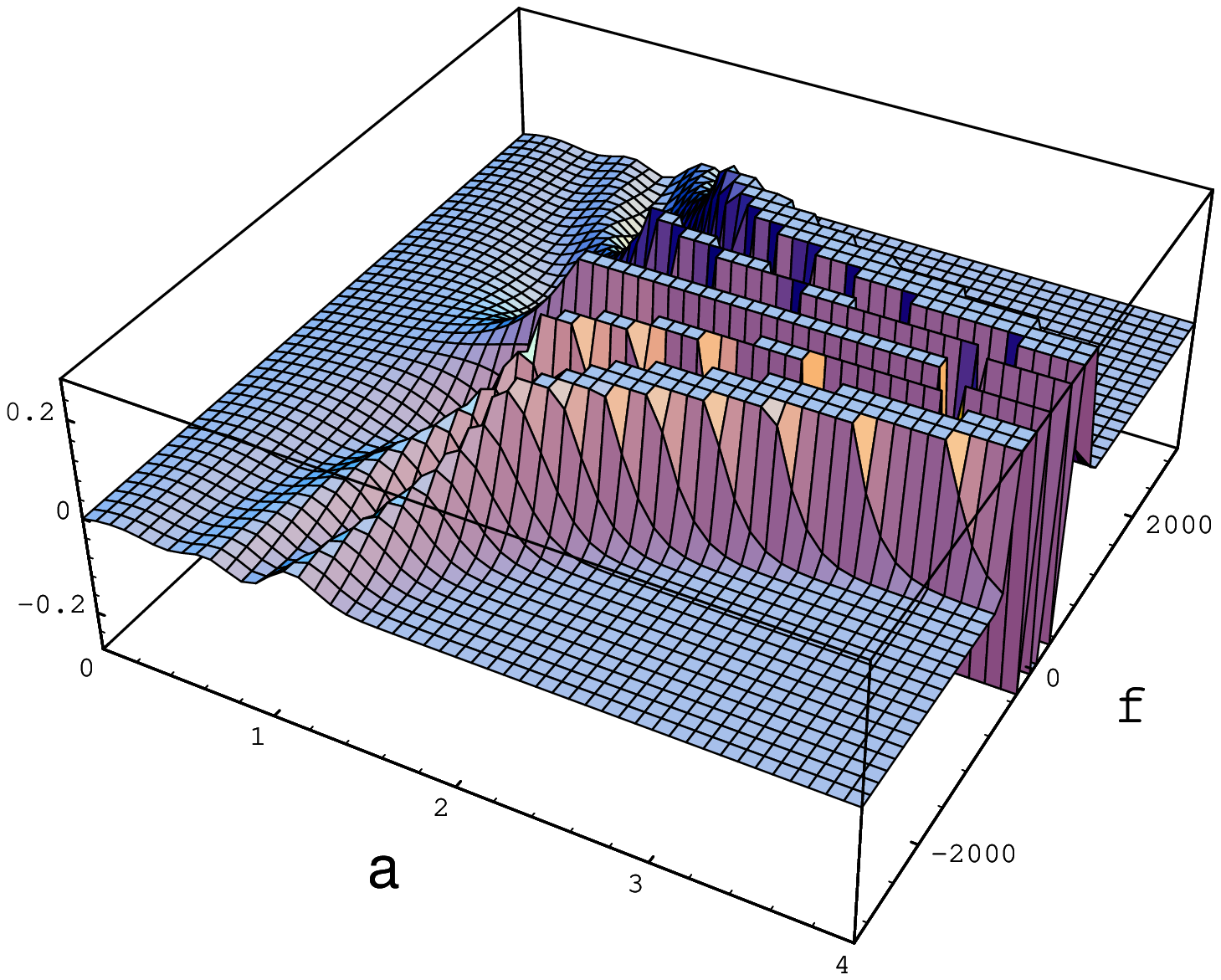}
\caption{Plot of $\psi_8:=\sum_{n=0}^8\sigma_n\Phi_n$, for
$\sigma_{\pm1}=\sigma_{-0}=0,~\sigma_0=1$ and $m=10^{-6}$.}
\end{figure}

\begin{figure}
\epsffile{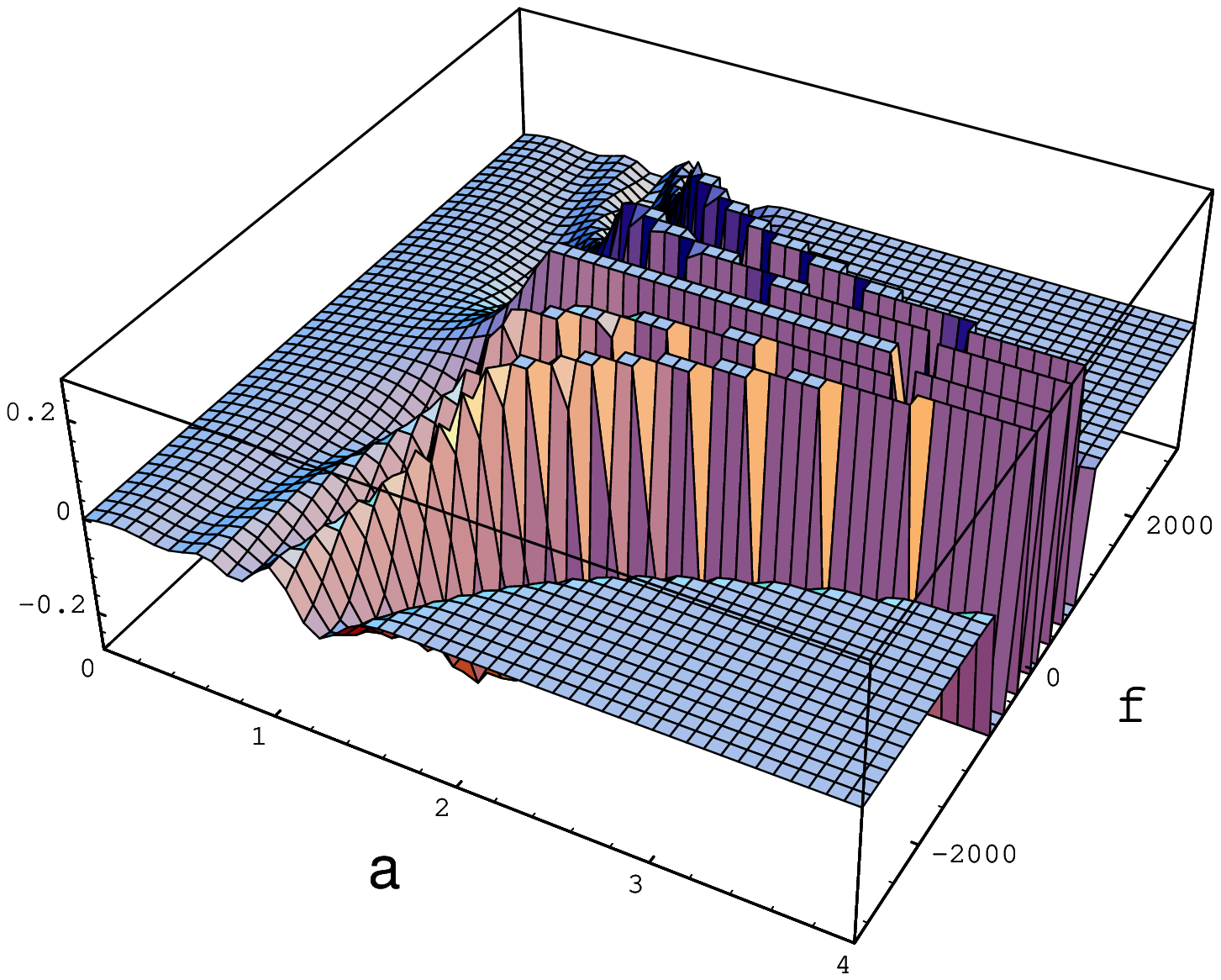}
\caption{Plot of $\psi_{10}:=\sum_{n=0}^{10}\sigma_n\Phi_n$, for
$\sigma_{\pm1}=\sigma_{-0}=0,~\sigma_0=1$ and $m=10^{-6}$.}
\end{figure}

\begin{figure}
\epsffile{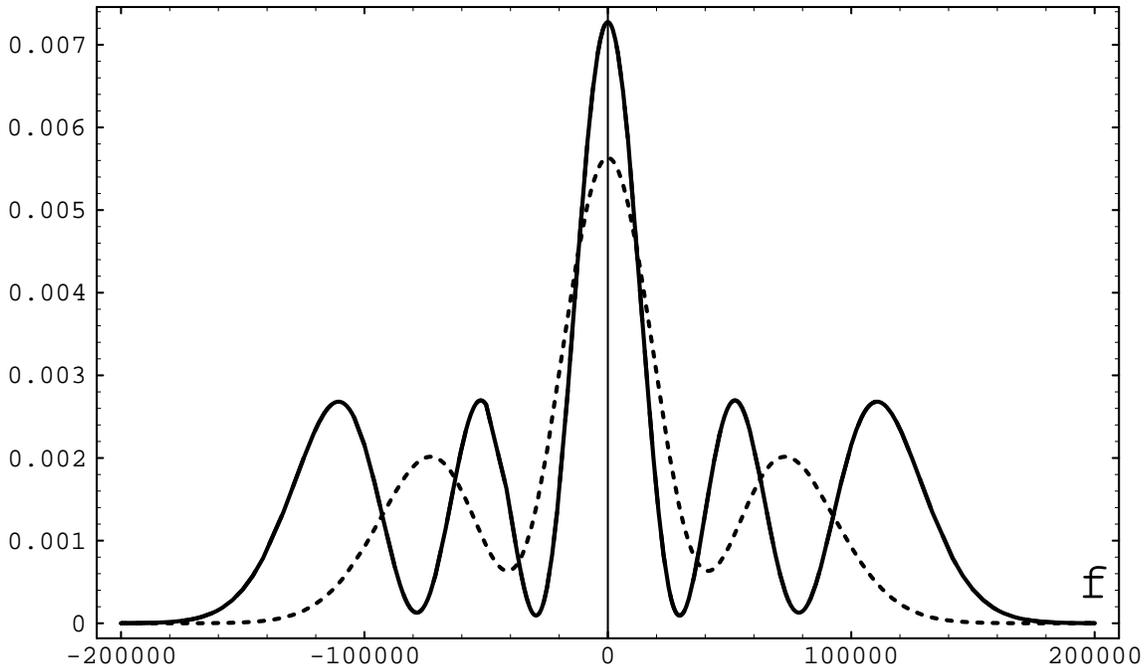}
\caption{Plot of the restriction of $\psi_6$ (dotted line),  $\psi_8$
(dashed line) and $\psi_{10}$ (full line) to the $a=0.1$ line. Note
that the graphs of $\psi_8$ and $\psi_{10}$ coincide for this
value of $a$. This shows the reliability of the finite mode
approximation. Note also the range of the values of
$\phi$ over which $\psi_n$ are non-zero.}
\end{figure}

\begin{figure}
\epsffile{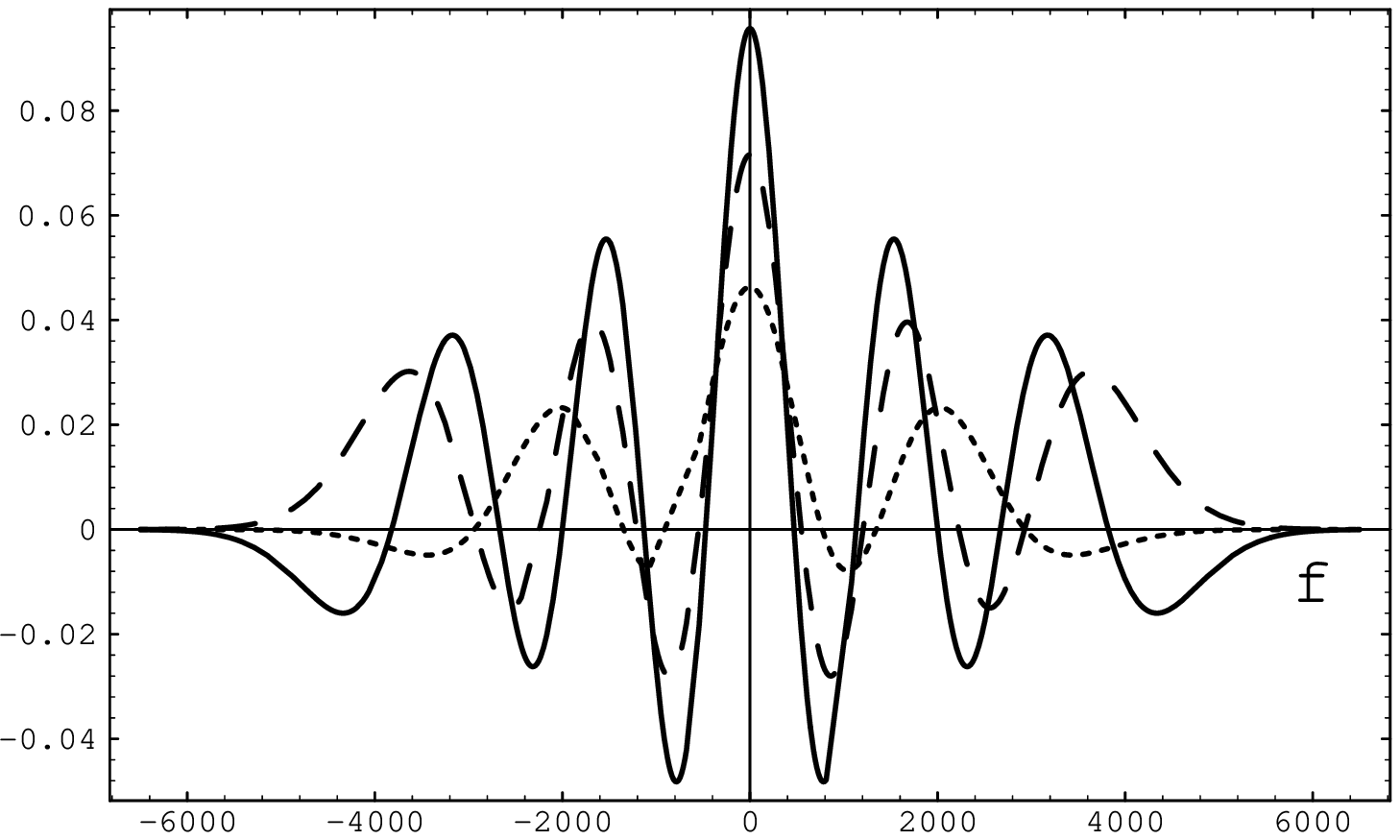}
\caption{Plot of the restriction of $\psi_6$ (dotted line),  $\psi_8$
(dashed line) and $\psi_{10}$ (full line) to the $a=1$ line. Note the
range of the values of $\phi$ over which $\psi_n$ are non-zero.}
\end{figure}

\begin{figure}
\epsffile{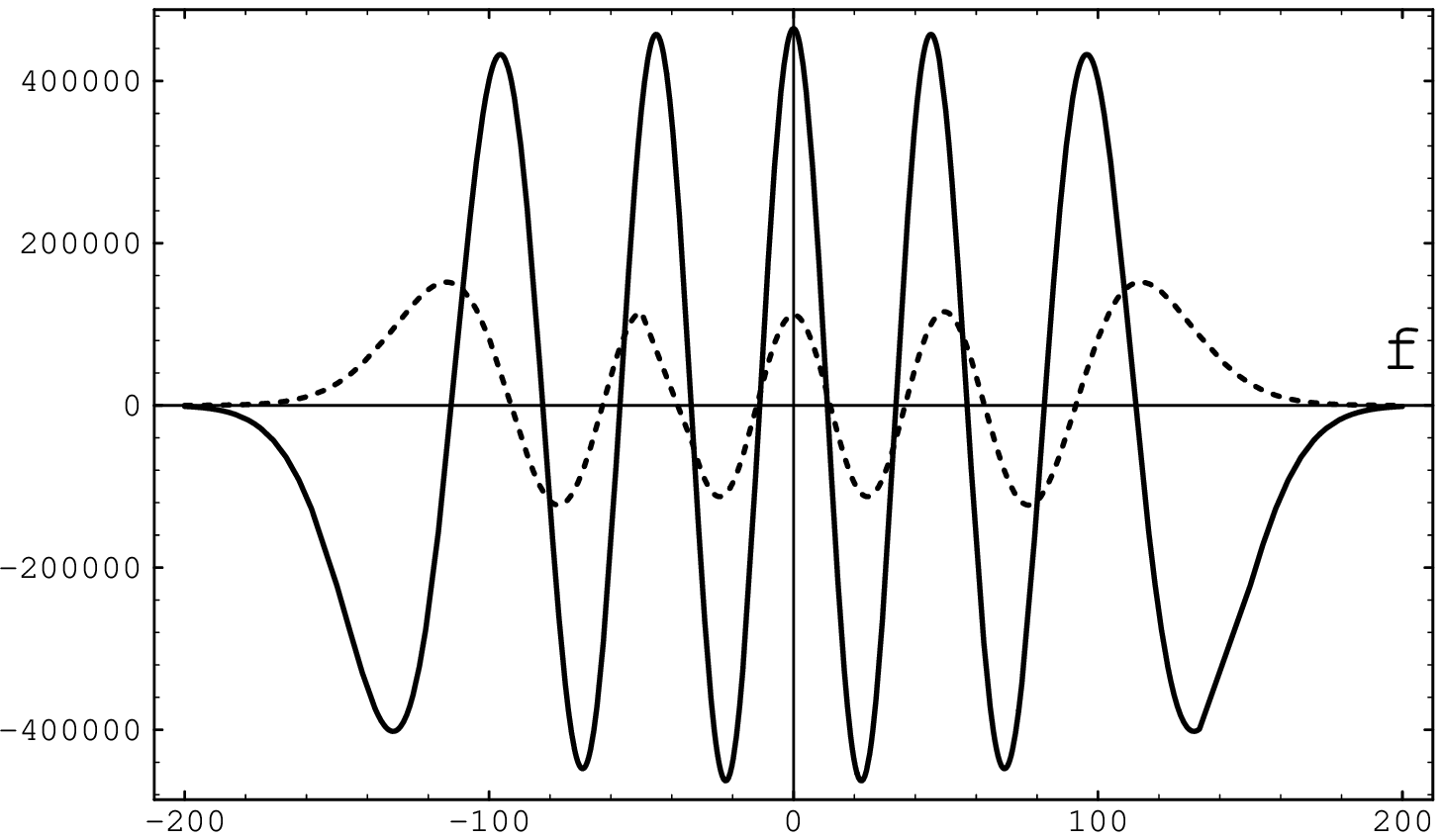}
\caption{Plot of the restriction of $\psi_8$ (dotted line), and $\psi_{10}$
(full line) to the $a=10$ line. Note the range of the values of $\phi$ over
which $\psi_n$ are non-zero.}
\end{figure}

\begin{figure}
\epsffile{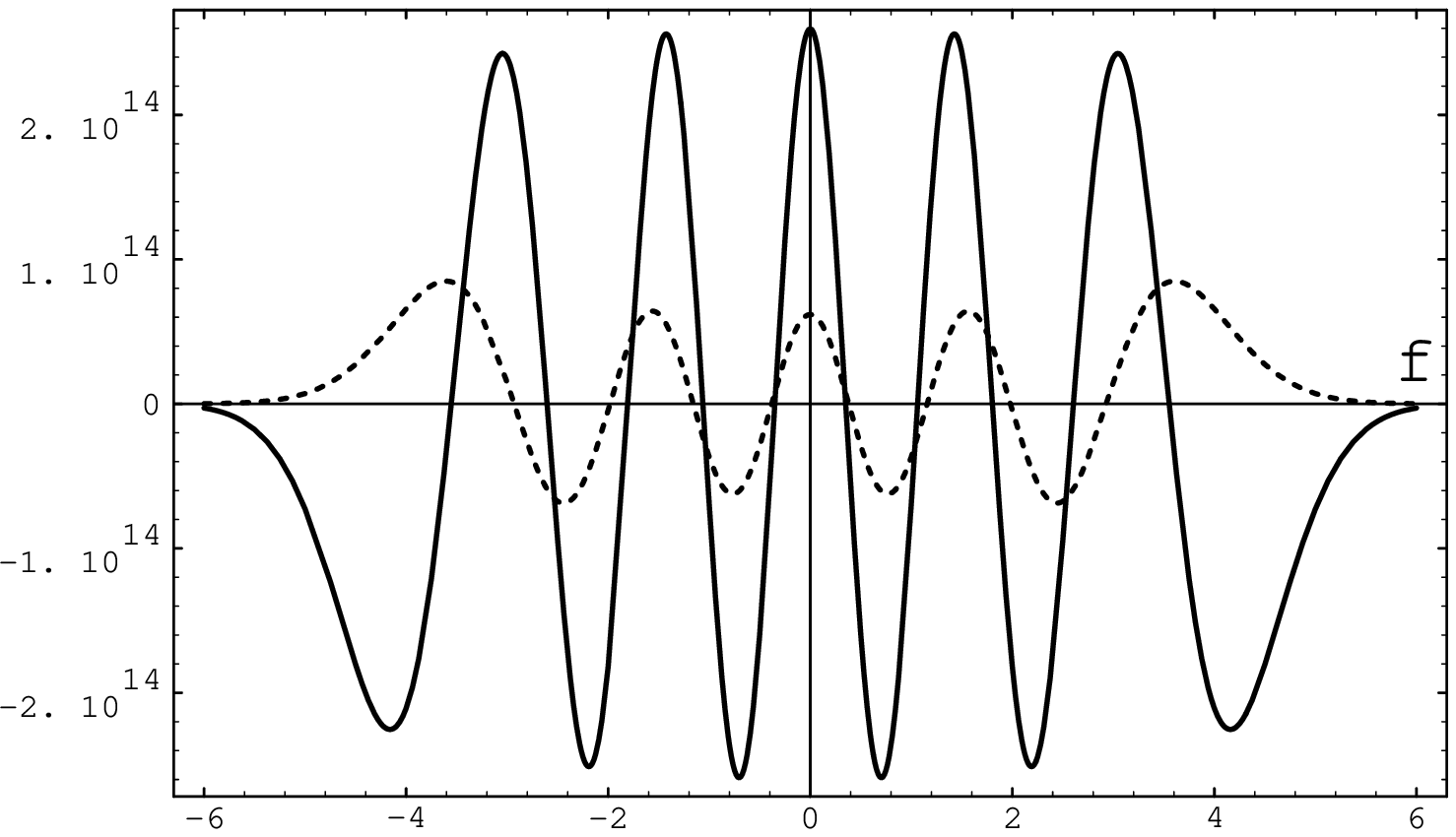}
\caption{Plot of the restriction of $\psi_8$ (dotted line), and $\psi_{10}$
(full line) to the $a=100$ line. Note the range of the values of $\phi$ over
which $\psi_n$ are non-zero.}
\end{figure}

\end{document}